\documentclass[aps,prl,twocolumn,groupedaddress]{revtex4-2}

\usepackage{graphicx}
\begin{document}


\title{Unusual Scaling of Kondo Spin Relaxation}


\author{Xingyu Shen and Yi Ji}
\email[]{yji@udel.edu}
\affiliation{Department of Physics and Astronomy, University of Delaware, Newark, Delaware 19716, USA}


\date{\today}

\begin{abstract}
The relation between the Kondo spin relaxation rate $ {\tau_{sK}}^{-1} $ and the Kondo momentum relaxation rate $ {\tau_{eK}}^{-1} $ is explored by using nonlocal spin valves with submicron copper channels that contain dilute iron impurities. A linear relation between $ {\tau_{sK}}^{-1} $  and $ {\tau_{eK}}^{-1} $  is established under varying temperatures. However, under varying impurity concentrations, ${\tau_{sK}}^{-1}$ remains nearly constant despite variation of $ {\tau_{eK}}^{-1} $  by a factor of 10. This surprising relation can be understood by considering spin relaxation through overlapping Kondo screening clouds and supports the physical existence of the elusive Kondo clouds.  
\end{abstract}


\maketitle


The Kondo effect~\cite{DeHaas36,Kondo64} has captured the attention of experimentalists and theorists alike for decades because of its complex many-body physics. In metals with dilute magnetic impurities, the experimental signature of Kondo effect is the low temperature increase of resistivity, which is attributed to the many-body antiferromagnetic s-d exchange interaction between the impurity spin and the conduction electron spins of the host metal. The Kondo effect has also been observed in semiconductor quantum dot (QD) systems where an unpaired spin in a QD is coupled to the surrounding electron reservoirs.~\cite{Goldhaber98} A popular but controversial physical picture of the Kondo effect is the Kondo screening cloud, which is an electron cloud surrounding the impurity site with an overall spin polarization opposite to the impurity spin. At temperatures well below the Kondo temperature $ T_K $, the net spin of the Kondo cloud completely screens the impurity spin forming a Kondo singlet state. The spatial extent $ \xi_K $ of the Kondo cloud is given by $ \hbar{v_F}/k_BT_K $  in ballistic transport regime and  $ \sqrt{{\hbar}D/k_{B}T_{K}} $ in diffusive regime,~\cite{Chandr00book,Affleck09} where $ v_F $ is the Fermi velocity, $ k_B $ is the Boltzmann constant, and $ D $ is the diffusion constant. Experimental evidence for the screening cloud is scarce and therefore its physical existence has been questioned.~\cite{Boyce74} Recently Borzenets $et~al.$~\cite{Borzenets20} found convincing evidence for micrometer-sized Kondo clouds in a QD system. In diffusive metals, $ \xi_K $ is expected to be $\sim 100$ nm, but has not yet been experimentally confirmed.

In recent years, the Kondo effect crosses paths with spintronics. In the Cu channels of nonlocal spin valves (NSLVs)~\cite{Johnson85,Jedema01} with dilute Fe impurities, the spin relaxation rate $ {\tau_{s}}^{-1} $ is found to increase at low temperatures complementing Kondo effect's low temperature increase of the momentum relaxation rate $ {\tau_{e}}^{-1} $.~\cite{Obrien14,Batley15,Hamaya16,Watts19} Here $ \tau_s $ and $ \tau_e $ are the spin relaxation time and momentum relaxation time, respectively. For spin relaxation in general, Elliott-Yafet (EY)~\cite{Elliott54,Yafet83} and Dyakonov-Perel (DP)~\cite{Dyakonov72} models give explicit relations between $ {\tau_s}^{-1} $  and $ {\tau_e}^{-1} $. The EY spin relaxation is caused by weak spin-orbit coupling between energy bands and $ {\tau_s}^{-1} $  is proportional to  $ {\tau_e}^{-1} $. The ratio $ \tau_e/\tau_s $ is the spin flip probability $ \alpha $.  The DP spin relaxation originates from spin-orbit coupling, caused by inversion symmetry breaking, between two spin subbands within the same energy band and the  $ {\tau_s}^{-1} $ is inversely proportional to $ {\tau_e}^{-1} $. The Kondo spin relaxation, however, is caused by s-d exchange interaction instead of spin orbit effects. The relation between the Kondo spin relaxation rate    $ {\tau_{sK}}^{-1} $ and Kondo momentum relaxation rate  $ {\tau_{eK}}^{-1} $, to the best of our knowledge,  has not yet been explored.

In this work, we extract values of $ {\tau_{sK}}^{-1} $ and $ {\tau_{eK}}^{-1} $ from Cu-based NLSVs fabricated by 2-step electron beam lithography. Each NLSV includes a spin injector $\mathrm{F_1}$, a spin detector $\mathrm{F_2}$, and a Cu channel, as shown in Figure 1 (a). Magnetic electrodes $\mathrm{F_1}$ and $\mathrm{F_2}$, made of $\mathrm{Ni_{81}Fe_{19}}$ alloy (permalloy or Py), are patterned in the first step and Cu channels are patterned in the second step. The materials are deposited by electron beam evaporation. Before the deposition of Cu, low energy ion milling is performed to clean the surface of Py and a 3 nm $\mathrm{AlO_x}$ layer is deposited. The Py/$\mathrm{AlO_x}$/Cu interface has been shown to provide a higher effective spin polarization than the ohmic Py/Cu interfaces.~\cite{Wang09,Cai16} The distance $L$ between $\mathrm{F_1}$ and $\mathrm{F_2}$ varies from 1 to 5  $\mu$m with 1 $ \mu $m increment. All Cu channels are 500 nm wide and 300 nm thick to prevent the suppression of Kondo clouds.~\cite{Chen91,Blachly95} This work involves data from two sample substrates (chip 11 and chip 12) with 10 devices on each. Devices on the same substrate undergo identical fabrication conditions. 

\begin{figure}
	\includegraphics[width=8.6cm]{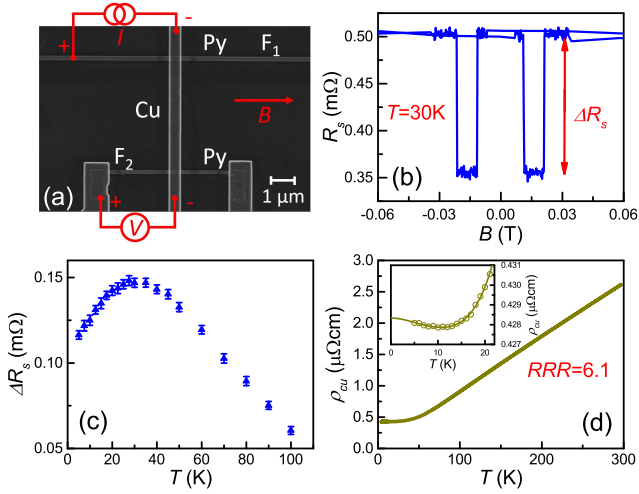}
	\caption{\label{fig1}(a) SEM image of a NLSV. Plots of (b) $R_s$ versus $B$, (c) $ \Delta{R}_s $ versus $T$, and (d) $ \rho_{cu} $ versus $T$ for device 11-43 ($ L=3.0\,\mu $m).}
\end{figure}

The measurement configuration is shown in Figure 1 (a). A low frequency excitation alternating current (AC) $ I_e $ is driven from $ \mathrm{F_1} $ to the upper end of the Cu channel, and the spin accumulation is detected by measuring the nonlocal voltage $V_{nl}$ between $ \mathrm{F_2} $ and the lower end of the channel. Figure 1 (b) shows the nonlocal resistance $ R_s={V_{nl}}/{I_e} $ as a function of magnetic field $B$ applied parallel to  $\mathrm{F_1}$ and $\mathrm{F_2}$ stripes. The high and low states of $R_s$ correspond to the parallel and antiparallel states of $\mathrm{F_1}$ and $\mathrm{F_2}$ magnetizations, respectively. The difference is the spin signal~\cite{Johnson93} 
\begin{equation} 
\Delta{R_s}=\frac{{P_e}^2\rho_{cu}\lambda_{cu}}{A_{cu}}e^{-\frac{L}{\lambda_{cu}}}, \label{Rs} 
\end{equation}
where $ P_e $ is the effective spin polarization of $\mathrm{F_1}$ and $\mathrm{F_2}$, $ \rho_{cu} $ the Cu resistivity, $ \lambda_{cu} $ the Cu spin relaxation length, and $ A_{cu} $ the Cu channel cross sectional area. $ \Delta{R_s}(T) $ of each NLSV is measured from 5 K to 100 K and Figure 1 (c) shows the data of device 11-43 (device 43 on chip 11). As $ T $ decreases, $ \Delta{R_s} $ initially increases, reaching its maximum at 30 K, and then decreases. This feature is well documented~\cite{Kimura08,Mihajlovic10,Zou12APL,Villamor13} for NLSVs and convincingly attributed to the Kondo effect.~\cite{Obrien14,Batley15,Hamaya16,Watts19,Obrien16,Kim17}

The resistivity $ \rho_{cu} $ of a given NLSV is deduced from its Cu channel resistance $ R_{cu} $, which is obtained by sending in a current through the channel and measuring the voltage difference between $ \mathrm{F_1} $ and $\mathrm{F_2}$. The $ \rho_{cu}(T) $ for device 11-43 is shown in Figure 1 (d) with $ \rho_{cu}=0.43 \,\mu\Omega\cdot $cm at 5 K and $ \rho_{cu}=2.60 \,\mu\Omega\cdot $cm at 295K. The ratio of the two values (6.1) is the residual resistivity ratio (RRR). The inset of Figure 1 (d) shows the low temperature portion of $ \rho_{cu}(T) $. The low $T$ increase of $ \rho_{cu} $ indicates Kondo effect from dilute magnetic impurities in Cu.  

Next, we extract the average $ P_e $ and $ \lambda_{cu} $ values of devices on the same substrate. $ \Delta{R_s} $ versus $ L $ is plotted for 10 devices on chip 11 at 30 K in Figure 2 (a). Fitting Eq.~(\ref{Rs}) to the plot yields $ \lambda_{cu}=2.6\pm0.1 \,\mu$m and $ P_e=0.066\pm0.003 $. The average  $ \rho_{cu} $ used in this process is deduced from the linear fitting of the $R_{cu}$ versus $L$ data in Figure 2 (b). In this manner, the average $P_e$ and $ \lambda_{cu} $ are obtained between 5 K and 100 K and shown in Figure 2 (c) and its inset, respectively. $ \lambda_{cu}(T) $ resembles $ \Delta{R_s}(T) $  in Figure 1(c) and reaches its maximum of 2.6 $ \mu $m at 30 K. $ \lambda_{cu} $ decreases to 2.2 $ \mu $m at 5 K because of the enhanced Kondo spin relaxation. The plot of $ P_e(T) $ shows a rather flat trend around 0.07 within the temperature range of our measurements. 

\begin{figure}
	\includegraphics[width=8.6cm]{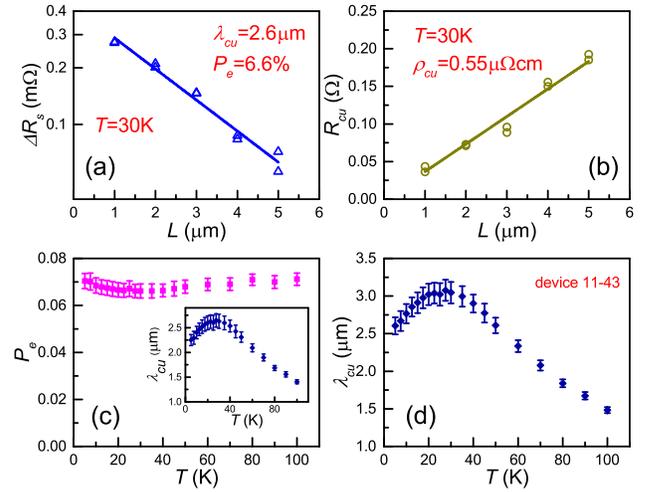}
	\caption{\label{fig2}(a) Spin signal $\Delta{R}_s$ and (b) Cu resistance $R_{cu}$ versus channel length $L$ for NLSVs on chip 11 at 30 K. (c) Fitted average $P_e$ and $ \lambda_{cu} $ (inset) as a function of $T$. (d) $ \lambda_{cu} $ versus $T$ for device 11-43. }
\end{figure}

As suggested by previous works on Py/Cu NLSVs, the Kondo effect originates from Fe impurities.~\cite{Obrien14,Batley15,Hamaya16,Obrien16,Kim17} The maximum $ \lambda_{cu} $ occurs at 30 K, which is the Kondo temperature $T_K$ for Fe impurities in Cu host. Data analysis of $ {\tau_s}^{-1}(T) $  and $ {\tau_e}^{-1}(T) $ later in the text is also consistent with $T_K = 30$ K. The Fe impurities are likely introduced in the fabrication processes. When the Py surface is ion milled, Fe atoms are removed and deposited on the side walls of the resist. When Cu is evaporated, the vapor flux of Cu transfers momentum to the Fe atoms on the side walls and redeposits them into the Cu channel. In some of the previous works,~\cite{Obrien14,Watts19,Obrien16,Kim17} Fe impurities are concentrated near the ohmic Py/Cu interfaces, and as a result the spin polarization $ P_e(T) $ is suppressed at low $T$. In our devices, the Fe impurities are located throughout the Cu channel. This is evident from the low $T$ upturn of $ \rho_{cu}(T) $, the low $T$ downturn of $ \lambda_{cu}(T) $, and the flat trend of $ P_e(T) $.

It is noticeable that data points disperse around the fitted lines in Figure 2 (a) and (b). For the two devices with $ L=3\,\mu $m, for example, data points of $ \Delta{R_s} $ are above the fitted line and those of $ R_{cu} $ are below. The two devices with $ L=4\,\mu $m have $ \Delta{R_s} $ below the fitted line and $ R_{cu} $ above. These indicate variations of $ \lambda_{cu} $ and $ \rho_{cu} $ between devices. Assuming a common $P_e$ (the fitted $ P_e $) for all devices on the same substrate at a specific $T$, we deduce $ \lambda_{cu} $ for each individual NLSV from its $ \Delta{R_s} $ and $ \rho_{cu} $ by using Eq.~(\ref{Rs}). $ \lambda_{cu}(T) $ for device 11-43 is shown in Figure 2 (d) with a maximum  $ \lambda_{cu}=3.0\pm0.1 \,\mu $m at 30 K. In this manner $ \lambda_{cu}(T) $ are obtained for all 20 NLSVs. The spin relaxation rate $ {\tau_s}^{-1}(T) $ is then calculated from $ \lambda_{cu}(T) $ by using the relation $ \lambda_{cu}=\sqrt{D\tau_s} $ and shown in Figure 3 (a) and (b) for devices 11-33 and 12-32, respectively. $ D=\frac{1}{3}{v_F}^2\tau_e $ is the diffusion constant and $ v_F=1.57\times10^6 $ m/s is the Fermi velocity of Cu. $ \tau_e $ can be derived from $ \rho_{cu} $ by using the Drude model $ \rho_{cu}=m/(\tau_ene^2) $, where $ n=8.47\times10^{28}\:\mathrm{m^{-3}} $ is the Cu electron density and $m$ and $e$ are electron mass and charge, respectively. With a decreasing $T$, $ {\tau_s}^{-1} $ initially decreases, reaches its minimum around 30 K, and then increases upon further cooling. This resembles Kondo effect’s low temperature increase of $ \rho_{cu} $ as shown in the insets of Figure 3 (a) and (b). The low $T$ increase of $ \rho_{cu} $ of 11-33 is much smaller than that of 12-32, indicating a lower impurity concentration in 11-33. However, the low $T$ increase of  $ {\tau_s}^{-1}$ of the two devices are surprisingly comparable. This provides the first hint for an unusual relation between Kondo momentum relaxation and Kondo spin relaxation. 

\begin{figure}
	\includegraphics[width=8.6cm]{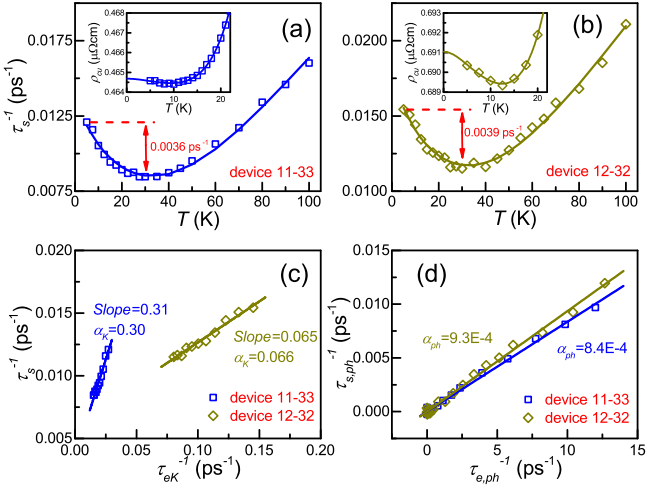}
	\caption{\label{fig3}Spin relaxation rate $ {\tau_s}^{-1} $  versus $T$ for (a) device 11-33 and (b) device 12-32. $ \rho_{cu}(T) $ plots are shown in the insets. (c) $ {\tau_s}^{-1} $  versus $ {\tau_{eK}}^{-1} $  for $ T \le 30 $ K for the two devices. The slopes of the linear fittings are compared with $ \alpha_K $ values obtained from fittings with Eq.~(\ref{taus}). (d) $ {\tau_{s,ph}}^{-1} $  versus $ {\tau_{e,ph}}^{-1} $  plots. }
\end{figure}

Applying Matthiessen's rule to spin relaxation, the total $ {\tau_s}^{-1} $ is given by $ {\tau_s}^{-1}={\tau_{s,def}}^{-1}+{\tau_{s,ph}}^{-1}+{\tau_{sK}}^{-1} $, where $ {\tau_{s,def}}^{-1} $, $ {\tau_{s,ph}}^{-1} $ , and $ {\tau_{sK}}^{-1} $  are the spin relaxation rates attributed to defects, phonon, and Kondo effects, respectively. Defining $ {\tau_{e,def}}^{-1} $, $ {\tau_{e,ph}}^{-1} $, and $ {\tau_{eK}}^{-1} $  as the corresponding momentum relaxation rates and $ \alpha_{def} $, $ \alpha_{ph} $, and $ \alpha_K $ as the associated spin flip probabilities, we have
\begin{equation}
	\frac{1}{\tau_s(T)}=\alpha_{def}\frac{1}{\tau_{e,def}}+\alpha_{ph}\frac{1}{\tau_{e,ph}(T)}+\alpha_{K}\frac{1}{\tau_{eK}(T)}. \label{taus}   
\end{equation}   
It is well justified to assume a linear relation between $ {\tau_s}^{-1} $  and  $ {\tau_e}^{-1} $ for defects and phonons, because EY mechanism is dominant in these processes. We will show later that $ {\tau_{sK}}^{-1} $  is also proportional to $ {\tau_{eK}}^{-1} $  under varying $T$.

The  $ {\tau_e}^{-1} $ of each type (total, defect, phonon, or Kondo) is linked to the corresponding $ \rho $ by the Drude model $ \rho=m/(\tau_ene^2) $. The defect resistivity $ \rho_{def} $ is $T$ independent and the phonon resistivity can be described as $ \rho_{ph}(T)=AT^5 $ at low $T$, where $A$ is constant related to the Debye temperature.~\cite{Zimanbook} The Kondo resistivity can be described by a phenomenological formula~\cite{Goldhaber98} 
\begin{equation}
	\rho_K(T)=\rho_{K0}{\left({\frac{{T_K'}^2}{T^2+{T_K'}^2}}\right)}^s, \label{Kondo}
\end{equation}
where $ T_K'=T_K/\sqrt{2^{1/s}-1} $, $s=0.225$ and $T_K = 30$ K. From $ {\tau_e}^{-1}={\tau_{e,def}}^{-1}+{\tau_{e,ph}}^{-1}+{\tau_{sK}}^{-1} $, the total resistivity is 
\begin{equation}
	\rho_{cu}(T)=\rho_{def}+AT^5+\rho_K(T). \label{rho}
\end{equation}
Fitting Eq.~(\ref{rho}) along with Eq.~(\ref{Kondo}) to the measured $ \rho_{cu}(T) $ data below 20 K yields $ \rho_{def} $, $A$, and $ \rho_{K0} $. Note that the fitting does not work well for $ T> $ 20 K, because $ \rho_{ph}(T)=AT^5 $ is an approximation valid at low $T$. For the data of 11-33 and 12-32 in the insets of Figure 3 (a) and (b), the fitted values of $ \rho_{K0} $ are 0.0013 $ \mu\Omega\cdot $cm and 0.0067 $ \mu\Omega\cdot $cm, respectively. $ \rho_{K0} $ or $ {\tau_{eK0}}^{-1} $  represents the $ \rho_K $ or $ {\tau_{eK}}^{-1} $ value at $T << T_K$.  

To extract $ \alpha_{def} $, $ \alpha_{ph} $, and $ \alpha_K $, we fit Eq.~(\ref{taus}) to the $ {\tau_s}^{-1}(T) $ data by using the empirical data of $ {\tau_{e,def}}^{-1} $, $ {\tau_{e,ph}}^{-1}(T) $, and $ {\tau_{eK}}^{-1}(T) $ obtained from the measured $ \rho_{cu}(T) $ and fitting. More specifically, $  {\tau_{e,def}}^{-1} $  can be obtained from the fitted $ \rho_{def} $ and $ {\tau_{eK}}^{-1}(T) $ from the fitted $ \rho_{K0} $ and Eq.~(\ref{Kondo}). For $ {\tau_{e,ph}}^{-1}(T) $  we use the relation $ \rho_{ph}(T)=\rho_{cu}(T)-\rho_{def}-\rho_K(T) $. We do not use $ \rho_{ph}(T)=AT^5 $ because it significantly deviates from experimental data when $ T>20 $ K. The best fits for $ \alpha_K $ are 0.30 $\pm$ 0.03 and 0.066 $\pm$ 0.006 and the best fits for $ \alpha_{ph} $ are $(8.4 \pm 0.3) \times 10^{-4}$ and $ (9.3\pm0.4) \times 10^{-4}$ for devices 11-33 and 12-32, respectively. While $ \alpha_{ph} $ values are comparable, $ \alpha_K $ values are quite different. Again, the results point to the unusual scaling for Kondo spin relaxation.

We should justify the assumed linear relation $ {\tau_{sK}}^{-1}(T)=\alpha_K\cdot{{\tau_{eK}}^{-1}(T)} $ under varying $T$ in Eq.~(\ref{taus}). In Figure 3 (c), $ {\tau_s}^{-1} $  is plotted versus $ {\tau_{eK}}^{-1} $   between 5 K and 30 K for the two NLSVs and we observe clear linear dependences. At $T\leq$ 30 K, the variation of $ {\tau_s}^{-1} $  should be dominated by $ {\tau_{sK}}^{-1} $ , because $ {\tau_{s,def}}^{-1} $  is $T$ independent and $ {\tau_{s,ph}}^{-1} $  is negligible compared to $ {\tau_{sK}}^{-1} $. Therefore, Figure 3 (c) confirms the linear relation between $ {\tau_{sK}}^{-1}(T) $ and $ {\tau_{eK}}^{-1}(T) $ under varying $T$. In addition, the slopes of the linear fittings to the $ {\tau_s}^{-1} $  versus $ {\tau_{eK}}^{-1} $  data are very close to the fitted $ \alpha_K $ values using Eq.~(\ref{taus}). Similarly, linear relation for phonons between $ {\tau_{s,ph}}^{-1}(T) $ and $ {\tau_{e,ph}}^{-1}(T) $ is also verified in Figure 3 (d). The data of $ {\tau_{s,ph}}^{-1} $  is obtained by subtracting $ \alpha_{def}\cdot{\tau_{e,def}}^{-1} $ and $ \alpha_{K}\cdot{\tau_{eK}}^{-1} $  from the total $ {\tau_s}^{-1} $. The slopes of the fitted lines are the same as the fitted $ \alpha_{ph} $ values by using Eq.~(\ref{taus}). 

\begin{figure}
	\includegraphics[width=8.6cm]{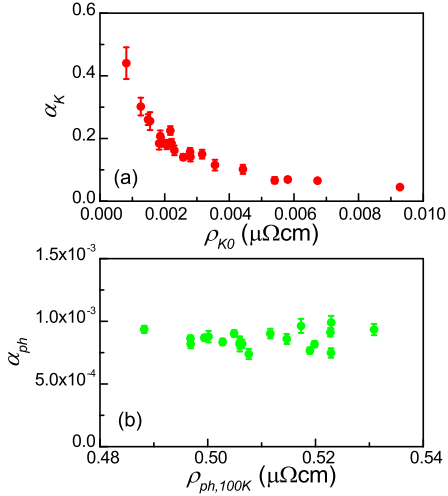}
	\caption{\label{fig4}(a) Kondo spin flip probability $ \alpha_K $ versus Kondo resistivity $ \rho_{K0} $. (b) Phonon spin flip probability $ \alpha_{ph} $ versus 100 K phonon resistivity $ \rho_{ph,100K} $.}
\end{figure}

Next, we demonstrate the unusual relation between $ {\tau_{sK}}^{-1} $  and $ {\tau_{eK}}^{-1} $ under a varying impurity concentration $ C_{Fe} $ which is approximately proportional to $ \rho_{K0} $ or $ {\tau_{eK0}}^{-1} $. Figure 4 (a) shows $ \alpha_K $ versus $ \rho_{K0} $ extracted from all NLSVs. Strikingly, $ \alpha_K $ decreases drastically from $ 0.44 \pm 0.05 $ to $ 0.045 \pm 0.004 $ as $ \rho_{K0} $ ($ \propto{\tau_{eK0}}^{-1} $) increases from $ < 0.001 \,\mu\Omega\cdot $cm to $> 0.009 \,\mu\Omega\cdot $cm. As a comparison, Figure 4 (b) shows $ \alpha_{ph} $ versus $ \rho_{ph,100K} $, which is the $ \rho_{ph} $ at 100K, for all NLSVs. $ \alpha_{ph} $ remains nearly a constant and independent of $ \rho_{ph,100K} $ as expected for processes governed by EY mechanism. The average value of $ \alpha_{ph} $ ($\sim 8.5\times10^{-4} $) is in good agreement with previous works.~\cite{Watts19,Villamor13,Monod79} The average value of $ \alpha_{def} $ is $ 3.2\times10^{-4} $ and the data are shown in the Supplementary Materials (Note S1). The decreasing trend in Figure 4 (a) suggests that the relation between $ {\tau_{sK0}}^{-1} $  and $ {\tau_{eK0}}^{-1} $  is not linear, where $ {\tau_{sK0}}^{-1} $ is the value of $ {\tau_{sK}}^{-1} $  at $T \ll T_K$. Figure 5 (a) shows $ {\tau_{sK0}}^{-1} $ , obtained by using the definition $ {\tau_{sK0}}^{-1}=\alpha_K\cdot{\tau_{eK0}}^{-1} $, versus $ {\tau_{eK0}}^{-1} $. While $ {\tau_{eK0}}^{-1} $  varies by a factor of 10,  $ {\tau_{sK0}}^{-1} $  stays nearly constant clearly defying a linear dependence. In contrast, the few previous theoretical treatments of Kondo spin relaxation assume a linear relation and yield a constant $ \alpha_K $ of $2/3$.~\cite{Kondo64,Kim17} The dependences shown in Figure 4 (a) and 5 (a) have been neither anticipated nor addressed previously. These plots with horizontal error bars are available in the Supplementary Materails (Note S2). The $ C_{Fe} $ for each NLSV can be extracted from the temperature $ T_{min} $ that corresponds to the minimum of the fitted $ \rho_{cu}(T) $ curve.~\cite{Obrien16,Franck61} Figure 5 (b) shows the extracted $ C_{Fe} $ versus $ \rho_{K0} $ for all NLSVs.

\begin{figure}
	\includegraphics[width=8.6cm]{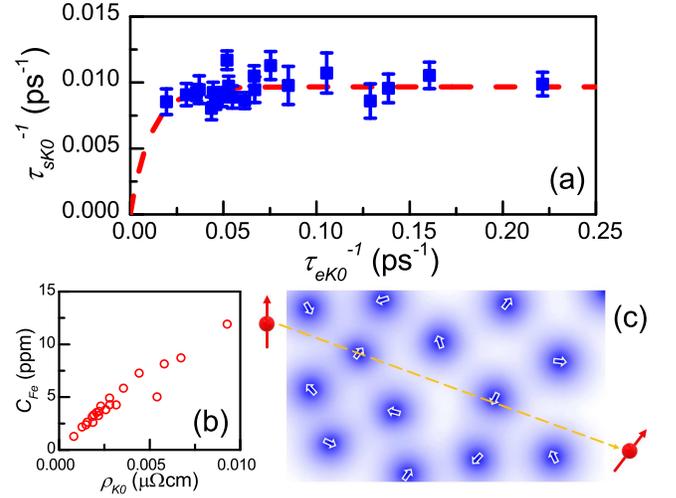}
	\caption{\label{fig5}(a) Kondo spin relaxation rate $ {\tau_{sK0}}^{-1} $  versus Kondo momentum relaxation rate $ {\tau_{eK0}}^{-1} $  from 20 NLSVs. (b) Fe impurity concentration $ C_{Fe} $ versus $ \rho_{K0} $. (c) Illustration of the Kondo medium. The gray scale indicates the spin density, and the white arrows indicate the polarization directions of the domains. }
\end{figure}

To address this unusual scaling between the Kondo momentum and spin relaxation, the physical picture of the Kondo cloud becomes appealing. If Kondo clouds exist, it is valid to consider them as momentum scattering barriers as well as spin scattering barriers for conduction electrons passing through them.~\cite{Simon03} The $ {\tau_{eK}}^{-1} $ should be proportional to the average charge density of the cloud. The $ {\tau_{sK}}^{-1} $ should be proportional to the average spin density of the cloud. It may also be related to the relative orientation between the conduction electron spin and the polarization direction of the cloud. The observed unusual scaling arises when the Kondo clouds of adjacent impurities overlap. 

Two relevant length scales are the size of a single Kondo cloud $ \xi_K $ and the average distance $ d_{Fe} $ between Fe impurities. The former is estimated to be $ \xi_K=\sqrt{{\hbar}D/k_BT_K}\approx 100 $ nm for diffusive Cu channels. The latter is 10 nm $ < d_{Fe} <$ 20 nm, estimated from the $ C_{Fe} $ of our NLSVs, and obviously $ \xi_K > d_{Fe} $. Therefore, the Kondo clouds from adjacent impurities overlap and the conduction electrons associated with the clouds form a continuous medium in the Cu channel. The medium can be characterized by its local charge density, spin density, and polarization direction with some spatial variations. The charge density of overlapping clouds should simply add up. However, the spin density of overlapping clouds may cancel out each other. Because impurity spin directions are random and so are the polarization directions of the clouds. Such cancellation effect of spin density has important implications on the $ {\tau_{sK}}^{-1} $. Figure 5 (c) is a qualitative illustration of the spin density distribution and polarization directions of the Kondo medium. Domains with random polarization directions are formed in the medium around impurity sites.

When a conduction electron traverses through the medium, the spin and momentum relaxation occur through the interaction between the electron and the Kondo medium. The $ {\tau_{eK0}}^{-1} $  or $ {\tau_{sK0}}^{-1} $  should be proportional to the average charge density or the average spin density of the medium, respectively, along the electron’s path. The influence of the polarization directions on $ {\tau_{sK0}}^{-1} $ can be neglected, because the traversing electron passes through many ($ \approx 10^4 $) randomly oriented Kondo domains within the time of $ \tau_{sK0} $. As $ C_{Fe} $ increases, more electrons are added to the Kondo medium, leading to a higher charge density and a higher $ {\tau_{eK0}}^{-1} $. However, the spin density may not increase, because a higher $ C_{Fe} $ enhances cloud overlapping and the cancellation effect. The exact trend is challenging to predict, because it requires precise knowledge of the spatial distributions of spin and charge densities of Kondo clouds and how overlapping clouds interact.  From experimental results in Figure 5 (a), we infer that the average spin density of the medium maintains a nearly constant value within the range of 1 ppm $ < C_{Fe} < $ 12 ppm, corresponding to 10 nm $ < d_{Fe} < $ 20 nm. The red curve in Figure 5 (a) is a guide to the eye with a reasonable assumption that $ {\tau_{sK0}}^{-1}\rightarrow 0 $ as $ {\tau_{eK0}}^{-1}\rightarrow 0 $. We  speculate that the initial slope of the curve, representing $ \alpha_K $ in the limit of $ {\tau_{eK0}}^{-1}\rightarrow 0 $, should be the theoretically predicted $ 2/3 $.~\cite{Kondo64,Kim17} 

In conclusion, we extract the Kondo momentum relaxation rate $ {\tau_{eK0}}^{-1} $  and the Kondo spin relaxation rate $ {\tau_{sK0}}^{-1} $  from Cu-based nonlocal spin valves with Fe impurities. While $ {\tau_{eK0}}^{-1} $  is tuned by a factor of 10 by varying Fe concentrations, $ {\tau_{sK0}}^{-1} $  remains nearly constant and defies a more intuitive linear dependence on $ {\tau_{eK0}}^{-1} $. Such a relation can be understood by considering a continuous Kondo medium formed by overlapping Kondo clouds. Spin relaxation occurs through interaction between a conduction electron spin and the medium. As the impurity concentration increases, the polarized spins of overlapping Kondo clouds partially cancel each other, and the average spin density of the Kondo medium reaches a stable value giving rise to a nearly constant $ {\tau_{sK0}}^{-1} $. Our experimental results provide evidence for the physical existence of the elusive Kondo screening clouds.

\subsection{}
\subsubsection{}

\bibliography{references}

\end{document}